\documentclass[aps,pra,longbibliography,showkeys,showpacs,groupedaddress,twocolumn,10pt]{revtex4-1}
\usepackage{amssymb}
\usepackage{graphicx}
\usepackage{dcolumn}
\usepackage{bm}
\usepackage{amsmath}
\usepackage{subfigure}
\usepackage{float}
\usepackage{multirow}
\usepackage{diagbox}
\usepackage{ulem}
\usepackage{color}
\usepackage{hhline}
\usepackage{color} %add color to a new command \fix

\begin{document}
\title{Cesium $nD_{J}$+$6S_{1/2}$ Rydberg molecules and their permanent electric dipole moments}

\author{Suying Bai\dag $^{1,2}$}
\author{Xiaoxuan Han\dag $^{1,2}$}
\author{Jingxu Bai$^{1,2}$}
\author{Yuechun Jiao$^{1,2}$}
\author{Jianming Zhao$^{1,2}$}
\thanks{Corresponding author: zhaojm@sxu.edu.cn, \dag These authors contributed equally to this work.}
\author{Suotang Jia$^{1,2}$}
\author{Georg Raithel$^{3}$}
\affiliation{$^{1}$State Key Laboratory of Quantum Optics and Quantum Optics Devices, Institute of Laser Spectroscopy, Shanxi University, Taiyuan 030006, China}
\affiliation{$^{2}$Collaborative Innovation Center of Extreme Optics, Shanxi University, Taiyuan 030006, China}
\affiliation{$^{3}$ Department of Physics, University of Michigan, Ann Arbor, Michigan 48109-1120, USA}
\date{\today}

\begin{abstract}
Cs$_2$ Rydberg-ground molecules consisting of a Rydberg, $nD_{J}$ (33 $\leq$ $n$ $\leq$ 39), and a ground state atom, 6$S_{1/2} (F=$3 or 4$)$, are investigated by photo-association spectroscopy in a cold atomic gas. We observe vibrational spectra that correspond to triplet $^T\Sigma$ and mixed $^{S,T}\Sigma$ molecular states. We establish scaling laws for the energies of the lowest vibrational states vs principal quantum number and obtain zero-energy singlet and triplet $s$-wave scattering lengths from experimental data and a Fermi model. Line broadening in electric fields reveals the permanent molecular electric-dipole moments; measured values agree well with calculations. We discuss the negative polarity of the dipole moments, which differs from previously
reported cases.
\end{abstract}
\maketitle

Recently, molecules formed between a ground-state and a Rydberg atom have attracted considerable attention due to their rich vibrational level structure and permanent electric dipole moments, which are unique for homonuclear molecules.
A Rydberg-ground molecule arises from low-energy scattering between the Rydberg electron and ground-state atoms located inside the Rydberg electron's wavefunction. This interaction, initially investigated in~\cite{Fermi,Omont}, has been predicted to lead to molecular binding in a novel type of Rydberg molecules, including the so-called trilobite~\cite{Greene} and butterfly molecule~\cite{Hamilton,Chibisov}.
The molecular bond length is on the order of the Rydberg-atom size
(a thousand Bohr radii a$_0$).
Rydberg-ground molecules were first reported in experiments with Rb~$nS_{1/2}$ ($n$ = 35-37) states~\cite{V. Bendkowsky} and later with Rb~$nP_{1/2,3/2}$~\cite{M. A. Bellos} and $nD_{3/2,5/2}$~\cite{D. A. Anderson, A. T. Krupp, Maclennan} states, as well as with Cs~$nS_{1/2}$~\cite{J. Tallant,Booth}, $nP_{3/2}$~\cite{HS} and $nD_{3/2}$~\cite{Shaffer2019} states.
The permanent electric dipole moment of S-type Rydberg-ground molecules has been measured to be $\thicksim$ 1~Debye for Rb~\cite{Li2011} and a few thousand Debye for Cs~\cite{Booth}.
The large size and the
permanent electric dipole moments of Rydberg-ground molecules
make these molecules good candidates for the realization of certain strongly correlated many-body gases~\cite{Weimer} and for quantum information processing~\cite{Lukin,Demille,Rabl}, as well as for dipolar quantum gases and spin systems with long-range interactions~\cite{Baranov,Kadau}.

Here we report on the measurement of long-range Cs$_2$ ($nD_{J}+6S_{1/2}F$) Rydberg-ground molecules for 33 $\leq$ $n$ $\leq$ 39, $J=3/2$ or 5/2, and $F=3$ or 4.
These molecules are deeply in the Hund's case(c)-regime, which differs from Rb $nD_{3/2,5/2}$-type molecules at lower $n$, which are Hund's case(a)~\cite{Maclennan}
or between Hund's case(a) and (c)~\cite{D. A. Anderson,A. T. Krupp}. Using a Fermi model, we calculate molecular potential energy curves (PECs), vibrational energies and
permanent electric-dipole moments.

The scattering interaction between the Rydberg electron and the ground-state atom is,
in the reference frame of the Rydberg ionic core~\cite{Omont},
\begin{equation}
\begin{aligned}
\widehat{V}(\textbf{r};R)=& 2\pi a_{s}(k)\delta^{3}(\textbf{r}-R\hat{\textbf{z}})\\
&+6\pi[a_{p}(k)]^{3}\delta^{3}(\textbf{r}-R\hat{\textbf{z}})\overleftarrow{\nabla}\cdot\overrightarrow{\nabla}
\end{aligned}
\end{equation}
where \textbf{r} and $R\hat{\textbf{z}}$ are the positions of the Rydberg electron and the perturber atom, $a_{l}(k)$ the scattering lengths, $k$ is the electron momentum, and $l$ the scattering partial-wave order (0 or 1 for $s$-wave or $p$-wave, respectively). The full Hamiltonian of the system is~\cite{Anderson},
\begin{equation}
\hat{H}(\textbf{r};R)= \hat{H}_{0}+\sum_{\substack{i=S,T}}\hat{V}(\textbf{r};R)\hat{P}(i)+A_{HFS}\hat{\textbf{S}}_{2}\cdot\hat{\textbf{I}}_{2}
\end{equation}
where $\hat{H}_{0}$ is the unperturbed Hamiltonian, which includes the spin-orbit interaction of the Rydberg atom. The second term sums over singlet ($i$ = $S$) and triplet ($i$ = $T$ ) scattering channels, using the projection operators $\hat{P}(T)=\hat{\textbf{S}}_{1}\cdot\hat{\textbf{S}}_{2}+3/4$, $\hat{P}(S)=1-\hat{P}(T)$ ($\hat{\textbf{S}}_{1}$ and $\hat{\textbf{S}}_{2}$ are the electronic spins of the Rydberg and ground-state atom, respectively). The last term represents the hyperfine coupling of $\hat{\textbf{S}}_{2}$ to the ground-state-atom nuclear spin $\hat{\textbf{I}}_{2}$, with hyperfine parameter $A_{HFS}$.
Numerical solutions of the Hamiltonian in Eq.~(2) on a grid of $R$-values yield sets of PECs. Figure~1(a) shows four PECs that are asymptotically connected with the atomic 36$D_{5/2}$-state. The PECs for $^{T}\Sigma$ for $6S_{1/2}F$=3 and $F$=4 are practically identical, while the PECs for $^{S,T}\Sigma$ are $\sim$ 10~MHz deeper for $F$=3 than
for $F$=4. A similar behavior was seen in Rb~\cite{D. A. Anderson, Maclennan} and Cs~\cite{HS}.

\begin{figure}[thb]
\vspace{-1ex}
\centering
\includegraphics[width=0.45\textwidth]{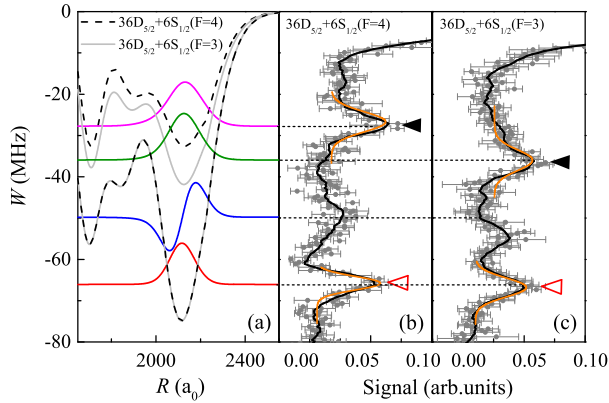}
\vspace{-1ex}
\caption{(a) PECs for $36D_{5/2}+6S_{1/2}(F=4)$ (dashed lines) and $36D_{5/2}+6S_{1/2}(F=3)$ molecules (gray solid lines), respectively. The deep potentials mostly arise from triplet $s$-wave scattering ($^{T}\Sigma$) and do not depend on $F$. The shallow potentials mostly arise from $s$-wave scattering of mixed $^{S,T}\Sigma$-states and depend on $F$; the PEC for $^{S,T}\Sigma$ $F=3$ is deeper than that for
$^{S,T}\Sigma$ $F=4$. The colored lines show the lowest vibrational wavefunctions on the PECs.
(b,c) Experimental photo-association spectra for $36D_{5/2}+6S_{1/2}(F=4)$ and $36D_{5/2}+6S_{1/2}(F=3)$ molecules. Energies are relative to the respective 36$D_{5/2}$ asymptotes. Filled (open) triangles mark the molecular signals formed by mixed $^{S,T}\Sigma$ (triplet $^{T}\Sigma$) potentials.  Gray symbols and error bars show data points, black lines display smoothed averages. The error bars are the standard error of ten independent measurements. The thin yellow lines display Gaussian fittings.}
\end{figure}

The experiment is performed in a crossed optical dipole trap (CODT) loaded from a magneto-optical trap (MOT). The CODT density, measured by absorption imaging, is $\thicksim 10^{11}$ cm$^{-3}$. This is sufficiently dense
to excite Rydberg-ground molecules with bond lengths $\sim 0.12~\mu$m (our case). After switching off the trapping lasers, two counter-propagated 852- and 510-nm lasers (pulse duration 3~$\mu$s) are applied to photo-associate the atoms into Rydberg-ground molecules. The lasers are both frequency-stabilized to the same high-finesse Fabry-Perot (FP) cavity
to less than 500~kHz linewidth. The 852-nm laser is 360~MHz blue-detuned from the intermediate $|6P_{3/2}, $F'$ = 5\rangle$ level.  The 510-nm laser is scanned from the atomic Rydberg line to $\sim$150~MHz below by scanning the radio-frequency signal (RF) applied to the electro-optic modulator  used to lock the laser to the FP cavity.  Rydberg molecules are formed when the detuning from the atomic line matches the binding energy of a molecular vibrational state. Rydberg atoms and molecules are detected using electric-field ionization and a microchannel plate (MCP) ion detector. Suitable timing of the MOT repumping laser allows us to prepare the atoms and molecules in either $F$=4 or $F$=3. The 510-nm laser can be tuned to excite either $nD_{5/2}+6S_{1/2}$ or $nD_{3/2}+6S_{1/2}$ molecules.

In Fig.~1 we show photo-association spectra of $36D_{5/2}+6S_{1/2}$ molecules for $F=4$ (Fig.~1(b)) and $F=3$ (Fig.~1(c)), respectively.
To reduce uncertainties, the spectra are averaged over ten measurements.
Both spectra display a pair of dominant molecular peaks, marked with triangles. They correspond to the vibrational ground ($\nu = 0$) states in the outermost wells of the shallow ($^{S,T}\Sigma$) and deep ($^{T}\Sigma$) PECs shown in Fig.~1(a), which arise from $s$-wave scatting. The deep, $^{T}\Sigma$ PEC corresponds with a triplet state of the Rydberg electron and the $6S_{1/2}$ atom. The two $^{S,T}\Sigma$ PECs correspond with mixed singlet-triplet states and have a reduced depth, which roughly is in proportion with the amount of triplet character in the molecular states. The binding energies of the $^{*}\Sigma , \nu=0$  states are extracted from Gaussian fits to the measured molecular peaks, with
statistical uncertainties on the order of 1~MHz.
Systematic uncertainties in the molecular line positions are negligible because of the high signal-to-noise ratio of the atomic reference lines in the spectra (relative to which the binding energies are measured), and because the FP cavity and the RF used to lock and scan the lasers have no significant drift.

\begin{figure}[htb]
\vspace{-1ex}
\centering
\includegraphics[width=0.45\textwidth]{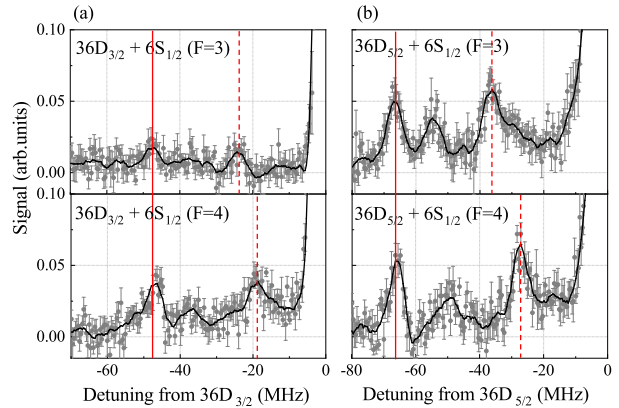}
\vspace{-1ex}
\caption{Measured spectra of $36D_{J}+6S_{1/2}$  molecules
for $J=3/2$ (a) and $J=5/2$ (b), for $F=3$ (top)
and $F=4$ (bottom). The laser detunings are relative to the atomic resonances, and the signal strengths
are displayed on identical scales.
Vertical solid and dashed lines mark the signals of the $^{T}\Sigma (\nu=0)$ and $^{S,T}\Sigma (\nu=0)$ ground vibrational states, respectively.
The signal strengths of the $J=5/2$ spectra are higher than those of the
$J=3/2$ ones, due to the higher excitation probability of the $nD_{5/2}$ atoms.}
\end{figure}

We have obtained the photo-association spectra
for all combinations of $J$ and $F$, for $n=33$ to 39.
In Fig.~2 we show the results for the case $n=36$. The $^{T}\Sigma, \nu =0$ and $^{S,T}\Sigma , \nu =0$ states are well-resolved and allow for accurate comparison of level energies between experiment and
theory. The $^{T}\Sigma$, $\nu =0$ levels, marked by solid vertical lines, do not depend on $F$.
Since the PECs for the measured states are largely due to $s$-wave scattering, the ratio of the binding energies of the $^{T}\Sigma$, $\nu =0$ levels between $J=3/2$ and $J=5/2$ is approximately given by the square of the ratio between the Clebsch-Gordan coefficients $\langle J, m_j=1/2 \vert m_\ell = 0, m_s = 1/2 \rangle$, with $J=3/2$ or $5/2$, and with magnetic quantum numbers $m_j$, $m_\ell$ and $m_s$ for the coupled, orbital and electron spins of the Rydberg electron, respectively. For $D$-type Rydberg-ground molecules in Hund's case (c), the binding-energy ratio is $\ell / (\ell + 1) = 2/3$, which is close to the binding-energy ratio evident in Fig.~2. The vertical dashed lines of Fig.~2 mark the $^{S,T}\Sigma, \nu=0$ states, which are mixed singlet-triplet. These are about half as deeply bound as $^{T}\Sigma, \nu=0$, whereby
$^{S,T}\Sigma, \nu=0$ for $F=3$ is about 5 to 10~MHz more deeply bound than $^{S,T}\Sigma, \nu=0$ for $F=4$.

For quantitative modeling of the singlet and triplet $s$-wave scattering length functions $a^T_{s}(k)$ and
$a^S_{s}(k)$, we have measured the binding energies of the states $^{T}\Sigma, \nu=0$ and $^{S,T}\Sigma, \nu=0$ for $nD_{5/2}+6S_{1/2}$ molecules with $n=33-39$, for both values of $F$.
 The measured data, listed in detail
in the Supplement,  are fitted with functions $a \, n^{*b_{Exp}}$, with effective quantum number $n^*$ and exponent $b_{Exp}$ (see Table I). The $b_{Exp.}$ are concentrated around $-5.60$, with one exception.
Calculated binding energies, listed in the Supplement, yield respective fitted exponents $b_{Theor.}$ that are within the uncertainty of the $b_{Exp.}$ (see Table I),
with the exception of the $^{S,T}\Sigma$ $F=4$ case, where the binding energies are smallest.
The $b$-values generally have a magnitude that is significantly less than $-6$. A value of $-6$ would be expected based on Rydberg wavefunction density. The deviation of $b$ from $-6$ may be attributed to the fact that at lower $n$ the molecules are less deep in Hund's case (c) than at higher $n$. This may diminish the binding of the $J=5/2$ molecules at lower $n$ and lead to a reduction of the magnitude of $b$. A modification of the scaling may also arise from $p$-wave-scattering-induced configuration mixing at lower $n$ as well as from the zero-point energy of the vibrational states.

The measured binding-energy data are employed to determine $s$-wave scattering lengths via comparison with model calculations similar to~\cite{Maclennan}.
The calculations yield best-fitting $s$-wave scattering-length functions for both singlet and triplet scattering, $a^S_{s}(k)$ and $a^T_{s}(k)$, with zero-energy scattering lengths $a^S_{s}(k=0)=-1.92$~a$_0$ and $a^T_{s}(k=0) = -19.16$~a$_0$;
a comparison with previous results is presented in the Supplement.
In our calculation we have included $p$-wave scattering
and found that it has only a small effect on the lowest vibrational resonances in the outermost wells of the PECs~\cite{suying2020}, within our $n$-range of interest. This is because the outermost wells are separated fairly well from further-in wells and are therefore strongly dominated by $s$-wave scattering, justifying our use of less accurate non-relativistic $p$-wave scattering-length functions $a^S_{p}(k)$ and $a^T_{p}(k)$~\cite{Khuskivadze2002}.

\begin{table}
 \caption{Fitted exponents $b$ (see text) for the scaling of the binding energies of $^{S,T}\Sigma \, \nu=0 $ and $^{T}\Sigma \, \nu=0$ states of $(nD_{5/2}+6S_{1/2} \, F)$ molecules, for $F=3$ and 4, over the range $33 \leqslant n \leqslant 39$. The fit function is $a \, n^{* b}$, with effective quantum number $n^*$ and exponent $b$.}
    \begin{center}
\begin{tabular}{|c|c|c|c|c|}
  \hline
  \multicolumn{1}{|c|}{\multirow{1}{*}{}} & S/T(F=3) & T(F=3) & S/T(F=4) & T(F=4)\\
  \hline
$b_{Exp.}$ & -5.65$\pm$ 0.38& -5.60$\pm$ 0.16& -6.19$\pm$ 0.14& -5.62$\pm$ 0.16\\
$b_{Theor.}$ & -5.68$\pm$ 0.01& -5.62$\pm$ 0.01& -5.55$\pm$ 0.01& -5.62$\pm$ 0.01\\
\hline
\end{tabular}
    \end{center}
\end{table}

\begin{figure}[htb]
\vspace{-1ex}
\centering
\includegraphics[width=0.45\textwidth]{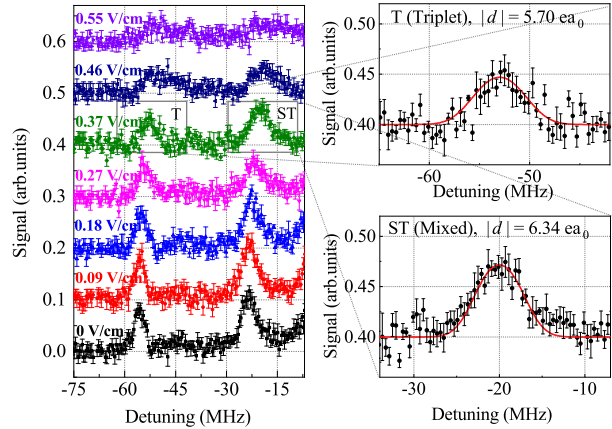}
\vspace{-1ex}
\caption{Spectra of 37$D_{5/2}+6S_{1/2}(F=4)$ Rydberg-ground molecules with indicated electric fields, $E$. The molecular peaks of $^{T}\Sigma, \nu=0$ and $^{S,T}\Sigma, \nu=0$ are blue-shifted by $E$ and substantially broadened in fields $E \geq 0.27$~V/cm. The right panel shows zoom-ins on the states $^{T}\Sigma(\nu=0)$ (top) and $^{S,T}\Sigma(\nu=0)$ (bottom). The red solid lines show model spectra based on Eq.~\ref{eq:lb} for dipole moments of magnitude $\vert d \vert =5.70~$ea$_0$ for  $^{T}\Sigma, \nu=0$ and 6.34~ea$_0$ for $^{S,T}\Sigma, \nu=0$, respectively.}
\end{figure}

Homonuclear Rydberg-ground molecules are unusual, in part, because of their permanent electric dipole moment, $d$, which are caused by configuration mixing. The values of $d$ are usually small in molecules with low-$\ell$ character, with the notable exception of Cs $S$-type molecules, where the quantum defect allows strong mixing with trilobite states~\cite{Booth}.
The values of $d_{i,\nu}$, with index $i$ denoting the PEC and $\nu$ the vibrational state, can be measured via
the broadening of the respective molecular line in an applied weak electric field, $E$. For electric-dipole energies, $-{\bf{d}}_{i, \nu} \cdot {\bf{E}}$, that are much smaller than the molecular binding energy, the line is inhomogeneously broadened about its center by a square function of full width $2 d_{i, \nu} E / h$ in frequency. This model applies if the moment of inertia of Rydberg molecules is very large and rotational structure cannot be resolved (our case).
The square function is convoluted with a Gaussian profile to  account for laser line broadening, electric-field inhomogeneities, magnetic fields etc. The standard deviation $\sigma_f$ of this Gaussian is experimentally determined by fitting field-free molecular lines.
The overall line profile, $S_{i, \nu}(\Delta f)$, as a function of detuning $\Delta f$ from the line center then is
\begin{equation}
\frac{h}{2 d E}  \left[   {\rm{erf}} \left( \frac{\Delta f + d_{i, \nu} E/h}{\sqrt{2} \sigma_f} \right)
                       -  {\rm{erf}} \left( \frac{\Delta f - d_{i, \nu} E/h}{\sqrt{2} \sigma_f} \right) \right].
                     \label{eq:lb}
\end{equation}
Since the field $E$ is accurately known from Rydberg Stark spectroscopy, the values of $\vert d_{i, \nu} \vert $ follow from comparing measured line shapes with profile functions calculated using Eq.~(3) over a range of test values for $\vert d_{i, \nu} \vert $.

In Fig.~3 we show line-broadening measurements
for $37D_{5/2}$+$6S_{1/2} (F=4)$ Rydberg molecules in several electric fields, as well as fit results based on
Eq.~\ref{eq:lb} for the vibrational ground states of
$^T \Sigma$ (top) and $^{S,T}\Sigma$ (bottom) PECs for the case $E=0.37$~V/cm. The
obtained dipole-moment magnitudes are 5.70 (6.34) $ea_0$ for the triplet (mixed) states. Analysis of the spectra for 0.18, 0.27 and 0.37~V/cm yields averaged dipole-moment magnitudes of 4.79 $\pm$ 0.78~$e a_0$ for $^T\Sigma$ and 5.49 $\pm$ 1.03~$e a_0$ for $^{S,T}\Sigma$.

For a comparison with theory, we first solve Eq.~(2) to obtain the PECs and electronic adiabatic dipole moments along the internuclear axis, $d_{i,z}(R)$. We then find the vibrational energies and wavefunctions, $\Psi_{i,\nu}(R)$, on the PECs~\cite{Anderson}. The dipole moments of the molecules, $d_{i,\nu}$, are
\begin{equation}
  d_{i,\nu} = \int\vert \Psi_{i,\nu}(R) \vert^2 d_{i,z}(R) dR .
\end{equation}
For the $^{T}\Sigma, \nu=0$ states we find $d_{i,\nu}$ values ranging between -4.85~ea$_0$ at $n=33$ and
-4.60~ea$_0$ at $n=38$. For $n=37$, the calculated dipole moment is -4.64~ea$_0$, which is in good agreement with the measured result ($|d|$ = 4.79 $\pm$ 0.78 $e a_0$).

We note that the molecular lines also exhibit a DC Stark shift due to the electric polarizability, $\alpha$, of the Rydberg atom. The atomic DC Stark shifts, $ - \alpha_{m_J} E^2 /2$, depend on the magnetic quantum number $m_J$ due to the tensor component of the polarizability. If the molecular Stark shift is less than the molecular binding, it can be calculated perturbatively as an average shift with weights $P(m_J)$, where $m_J$ is in the laboratory frame (defined by the direction of the applied electric field). Figure~3 further includes a hint that the molecular lines may split in stronger electric fields (top curve for 0.55~V/cm). The DC Stark shifts and possible splittings can result in an overestimate of the molecular dipole moment; this may explain the deviations between measured and calculated dipole moments.

\begin{figure}[htb]
%\vspace{-1ex}
\centering
\includegraphics[width=0.5\textwidth]{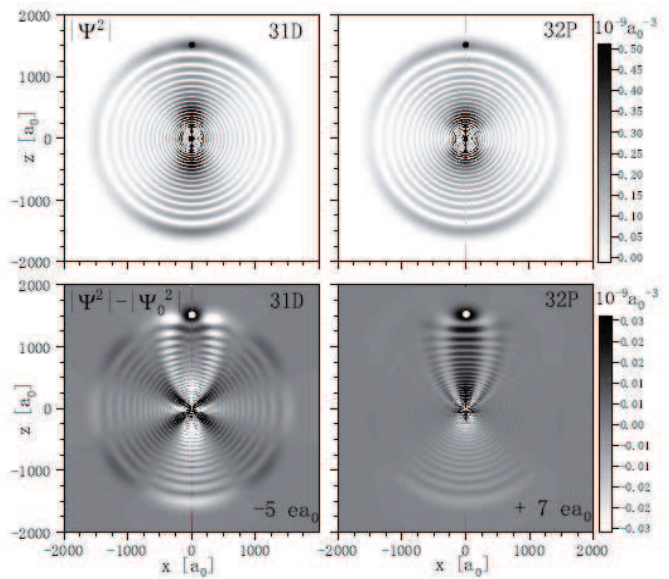}
%\vspace{1ex}
\caption{Densities of adiabatic electronic wavefunctions for Cs $31D_{5/2}$ + $6S_{1/2}$ $(F=4)$ $^T \Sigma$ (left) and $32P_{3/2}$ + $6S_{1/2}$ $(F=4)$ $^T \Sigma$ (right panels), with the perturber located at $\approx$ 1500$a_0$ (dot). Top: wavefunction densities. Bottom: difference between electronic wavefunction densities of molecules and atoms on a linear gray-scale, with white and black indicating reductions and increases by
amounts shown on the gray-scale bar. 
The $P$-type molecular state (right) carries a trilobite-like component that interferes mostly constructively with the $P$-orbital, causing a positive dipole moment of about 7~ea$_0$. In the case of the $D$-type molecule (left), the trilobite orbital predominantly shows destructive interference with the $D$-orbital, causing a negative dipole moment of about -5~ea$_0$. }
\end{figure}

While the current measurement method does not give the sign of $d_{i,\nu}$, the calculations reveal that the $d_{i,\nu}$ of Cs~$nD_{J}$-type Rydberg-ground molecules are \textit{negative}, which differs from reports on other types of Rydberg-ground molecules~\cite{J. Tallant,Booth,HS,Markson}. Physically, the sign of $d_{i,\nu}$ reflects the direction of the electronic charge shift along the axis of the Rydberg molecule relative to the location of the Cs~6$S_{1/2}$ atom. The direction of the weak electric field $E$ applied to measure the dipole moment is not relevant, as long as the field is weak (our case). A negative
$d_{i,\nu}$ corresponds with a deficiency of electron charge from the vicinity of the Cs~6$S_{1/2}$ perturber atom. This situation can generally be described as destructive interference of the Rydberg electron wavefunction near the perturber or, equivalently, as a possible case of electronic configuration mixing near the perturber (LCAO picture).

For further illustration, in Fig.~4 we show electronic wavefunctions of Cs $D$-type and $P$-type Rydberg-ground molecules in the outer well of the respective PECs (see Fig.~1 for typical PECs). Since the configuration mixing is weak, in the bottom panels in Fig.~4 we plot the difference of the wavefunction density relative to that of the unperturbed atomic state. An analysis of the electronic states by $\ell$- and $m$-quantum numbers shows that the $D$-type molecule mostly mixes with $P$ orbitals and with a combination of high$-\ell$ states similar to the trilobite state~\cite{Greene}, while the $P$-type molecule mostly mixes with $D$ orbitals and the trilobite-like state. Admixtures from $S$- and $F$-states are smaller. The admixture probabilities $\sim 10^{-4}$, corresponding to a typical wavefunction density variation on the order of a few percent, as seen in Fig.~4, leading to $\vert d_{i,\nu} \vert$-values much smaller than the wavefunction diameter. In Fig.~4 it is seen that the $P$-state molecule exhibits predominantly constructive interference near the perturber, corresponding to a positive dipole moment. A similar mixing analysis was reported for Rb~($35S$+$5S$) molecules with a small positive dipole moment~\cite{Li2011}. Interestingly, for the $D$-state molecule in Cs the mixing near the perturber is predominantly destructive, corresponding to a negative dipole moment.

In summary, we have observed Cs~$nD$ Rydberg-ground molecules involving Rydberg-state fine structure and ground-state hyperfine structure. Measurements of the binding energies for $^{T}\Sigma (\nu=0)$ and $^{S,T}\Sigma (\nu=0)$ molecular vibrational states were modeled with calculations. We have measured permanent electric dipole moments with magnitudes of a few ~ea$_0$. Calculations
show that the dipole moment is \textit{negative}. Future work may further elucidate this behavior, the exact shifts and splittings due to the tensor atomic polarizability, as well as the transition from weak to large
electric-dipole energy shifts relative to the molecular binding.

The work was supported by the National Key R$\&$D Program of China (Grant No. 2017YFA0304203), the National Natural Science Foundation of China (Grants Nos. 11434007, 61835007, 61675123, 61775124 and 11804202), Changjiang Scholars and Innovative Research Team in University of Ministry of Education of China (Grant No. IRT\_17R70) and 111 project (Grant No. D18001) and 1331KSC.

\end{document}